\documentclass[pdflatex,sn-mathphys-num]{sn-jnl}% Math and Physical Sciences Numbered Reference Style 
\usepackage{graphicx}%
\usepackage{multirow}%
\usepackage{amsmath,amssymb,amsfonts}%
\usepackage{amsthm}%
\usepackage{mathrsfs}%
\usepackage[title]{appendix}%
\usepackage{xcolor}%
\usepackage{textcomp}%
\usepackage{manyfoot}%
\usepackage{booktabs}%
\usepackage{algorithm}%
\usepackage{algorithmicx}%
\usepackage{algpseudocode}%
\usepackage{listings}%
\usepackage{adjustbox}
\theoremstyle{thmstyleone}%
\theoremstyle{thmstyletwo}%

\theoremstyle{thmstylethree}%

\begin{document}

\title[Discovering Boundary Equations for Wave Breaking using Machine Learning]{Discovering Boundary Equations for Wave Breaking using Machine Learning}


\author*[1]{\fnm{Tianning} \sur{Tang}}\email{tianning.tang@eng.ox.ac.uk}

\author*[2]{\fnm{Yuntian} \sur{Chen}}\email{ychen@eitech.edu.cn}

\author[3]{\fnm{Rui} \sur{Cao}}\email{rui.cao17@imperial.ac.uk}

\author[1]{\fnm{Wouter} \sur{Mostert}}\email{wouter.mostert@eng.ox.ac.uk}

\author[4]{\fnm{Paul H.} \sur{Taylor}}\email{paul.taylor@uwa.edu.au}

\author[1]{\fnm{Mark L.} \sur{McAllister}}\email{mmc@woodthilsted.com}

\author[5]{\fnm{Bing} \sur{Tai}}\email{taibing@just.edu.cn}

\author[6]{\fnm{Yuxiang} \sur{Ma}}\email{yuxma@dlut.edu.cn}

\author[3]{\fnm{Adrian H.} \sur{Callaghan}}\email{a.callaghan@imperial.ac.uk}

\author[1]{\fnm{Thomas A. A.} \sur{Adcock}}\email{thomas.adcock@eng.ox.ac.uk}

\affil[1]{\orgdiv{Department of Engineering Science}, \orgname{University of Oxford}, \orgaddress{\street{Parks Road}, \city{Oxford}, \postcode{OX1 3PJ}, \country{United Kingdom}}}

\affil*[2]{\orgname{Eastern Institute for Advanced Study}, \orgaddress{\city{Ningbo}, \postcode{315000}, \state{Zhejiang}, \country{China}}}

\affil[3]{\orgdiv{Department of Civil and Environmental Engineering}, \orgname{Imperial College London}, \orgaddress{\street{Skempton Building}, \city{London}, \postcode{SW7 2AZ}, \country{United Kingdom}}}

\affil[4]{\orgdiv{School of Earth and Oceans}, \orgname{The University of Western Australia}, \orgaddress{\street{35 Stirling Highway}, \city{Perth}, \postcode{WA 6009}, \country{Australia}}}

\affil[5]{\orgdiv{School of Naval Architecture and Ocean Engineering}, \orgname{Jiangsu University of Science and Technology}, \orgaddress{\city{Zhenjiang }, \postcode{212100}, \country{China}}}

\affil[6]{\orgdiv{State Key Laboratory of Coastal and Offshore Engineering}, \orgname{Dalian University of Technology}, \orgaddress{\city{Dalian}, \postcode{116023}, \country{China}}}


\abstract{Many supervised machine learning methods have revolutionised the empirical modelling of complex systems. These empirical models, however, are usually "black boxes" and provide only limited physical explanations about the underlying systems. Instead, so-called “knowledge discovery” methods can be used to explore the governing equations that describe observed phenomena. This paper focuses on how we can use such methods to explore underlying physics and also model a commonly observed yet not fully understood phenomenon -- the breaking of ocean waves.
In our work, we use symbolic regression to explore the equation that describes wave-breaking evolution from a dataset of in silico waves generated using expensive numerical methods. Our work discovers a new boundary equation that provides a reduced-order description of how the surface elevation (i.e., the water-air interface) evolves forward in time, including the instances when the wave breaks -- a problem that has defied traditional approaches. Compared to the existing empirical models, the unique equation-based nature of our model allows further mathematical interpretation, which provides an opportunity to explore the fundamentals of breaking waves. Further expert-AI collaborative research reveals the physical meaning of each term of the discovered equation, which suggests a new characteristic of breaking waves in deep water -- a decoupling between the water-air interface and the fluid velocities. This novel reduced-order model also hints at computationally efficient ways to simulate breaking waves for engineering applications.}




\maketitle

\section{Introduction}
Wave breaking is a process familiar to many but about which significant scientific questions remain \cite{cokelet1977breaking,thorpe1980bubbles,deconinck2023dominant,lamarre1991air,melville2002distribution,mcallister2024three}. The detailed physics and statistics of wave breaking is an open question despite its scientific importance in engineering \cite{longuet1980unsolved,bullock2007violent,bredmose2010breaking} and oceanography \cite{banner1993wave,Melville1996,babanin2011breaking,deike2022mass,hafner2023machine,ardhuin2010semiempirical,Callaghan2024}. Despite its importance and being commonplace, modelling breaking waves still presents significant challenges due to the complex air-water behaviour during wave breaking. To explicitly resolve turbulence and capture the full wave evolution, high-fidelity numerical simulations are required, which solve the Navier–Stokes equations directly but at a very high computational cost \cite{tian2012eddy,deike2015capillary,mostert2022high}. In the Eulerian potential flow framework, however, the wave-breaking effects need to be artificially introduced with some pre-assumed functional form and empirical fitted coefficients. A few recent studies have taken tentative steps to apply machine learning to the problem \cite{eeltink2022nonlinear,liu2024machine}. Despite these models having shown reasonable performances with many highly tuned parameter values, very few physical insights into the fundamentals of breaking waves have emerged from these works.

To address this, we take a more radical machine-learning approach with symbolic regression. We aim to discover a new mathematical description for the full evolution of a breaking wave without any pre-assumed functional forms. We train our model using high fidelity Volume of Fluid (VoF) simulations of waves, validate the results against independent experiments and numerical results using other models, and compare with well-established analytical results. We aim to find a validated mathematical equation that describes full wave evolution during breaking as an extension to the standard model used for non-breaking waves. Additionally, the discovered equation can be further interpreted within the context of existing wave theory, which allows for further mathematical exploration of the fundamentals of breaking wave phenomenon. 

We see this as an example of ``knowledge discovery''  with machine learning, a novel extension to the scientific method which represents a deep integration of domain expertise with interpretable machine learning, enhancing human understanding of physical principles through expert-AI interaction. It marks an advancement in scientific paradigms, signifying a progression from black-box data-driven models based on experimental or simulation data to AI-assisted and AI-inspired scientific research focused on human-comprehensible mechanisms.

\begin{figure*}%[tbhp]
\centering
\includegraphics[width=1\linewidth]{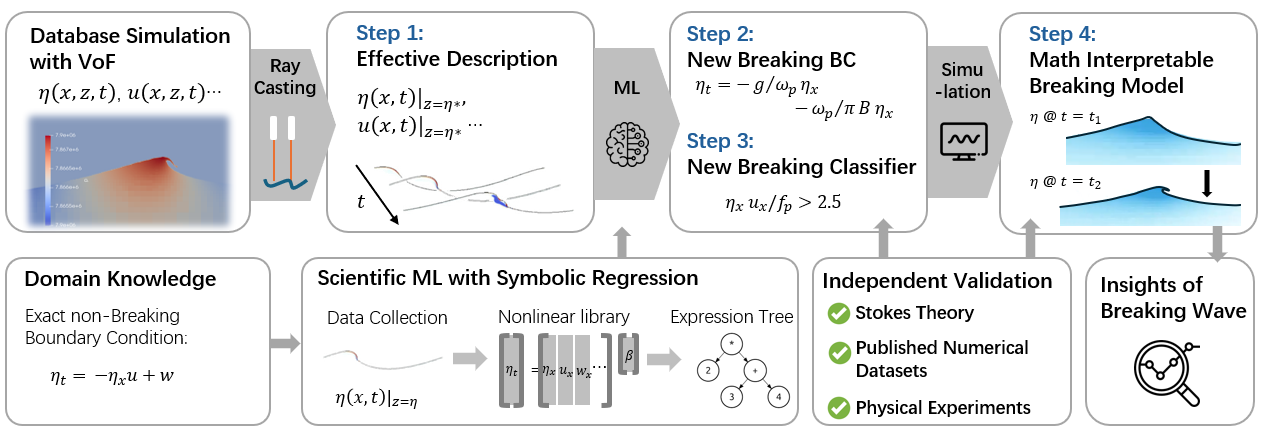}
\caption{\textbf{Overview of our study.} We start with large amounts of high-fidelity numerical simulated data. We then describe the air-water interface with a ray-casting approach. We use the domain knowledge from non-breaking waves to inform the construction of the nonlinear library, where in this study we focused on the evolution of the air-water interface, so only the kinematics boundary condition is considered. We then use symbolic regression and classification to discover a novel equation that models the evolution of the air-water interface during wave breaking, and also an expression that classifies the spatio-temporal location of the breaking region. We combine the breaking Boundary Condition (BC) and the classifier into an operational wave-breaking model, where we further analyse the performance on unseen experimental, independent datasets and compare them to existing theory. Finally, we reveal the physical insights from the validated breaking wave model in a human-AI collaborative approach.}
\label{fig:map}
\end{figure*}

We present our paper as follows. We first demonstrate the practicability of using such an interpretable machine learning approach for wave evolution via re-discovering the existing boundary equations for non-breaking waves (Section \ref{sec:redis}). We then present our approach to derive our new breaking wave model as illustrated in Fig \ref{fig:map} through the following steps:

\begin{enumerate}
    \item A new description of Boundary Condition (BC) for overturning waves that avoids non-single values of surface elevation in an Eulerian specification (Section \ref{sec:raycast}).
    \item Discovering a novel kinematic boundary equation for breaking wave evolution via symbolic regression (Section \ref{sec:ke}).
    \item Learning a new expression to classify the breaking region (Section \ref{sec:class}).
    \item Constructing an approximate breaking model by applying the new breaking boundary equation within the identified breaking region (Section \ref{sec:sim}).
\end{enumerate}

 %Having developed our approach, in addition to the standard unseen test cases, we further analyse the performance of the discovered equation with three additional independent sources: the theoretical geometric breaking criterion following Stokes expansion, a published numerical wave-breaking dataset with a different underlying governing equation, and an independent physical wave experiment. 
 After validating our new breaking model on a wide range of datasets, we finally discuss the physical implication of our results (Section \ref{sec:math}).

\section{Re-discovering the Fully Nonlinear Boundary Condition (FNBC) for non-breaking Waves}
\label{sec:redis}
We first examine the practicability of applying such a symbolic regression approach to find the wave evolution equations by re-discovering the fully nonlinear boundary condition (FNBC) for non-breaking waves. This non-breaking BC is well known and is used in for fully nonlinear potential flow theory \cite{holthuijsen2010waves}: 
\begin{equation}
    \eta_{t,\text{FNBC}} = - \eta_x u+  w,
    \label{eqn:fnpf}
\end{equation}
where $\eta$ is the surface elevation, subscript $_t$ and $_x$ denote the temporal and spatial derivatives, $u$ is the horizontal fluid velocity at the surface and $w$ is the vertical fluid velocity at the surface. The waves propagate in the positive x-direction and $z$, $w$ and $\eta$ are positive upwards.

\subsection{Non-breaking Wave Database}
We first establish a spatial-temporal wave evolution database for non-breaking waves in deep water. This database is obtained through numerical simulation of unidirectional focused wave groups with a fully nonlinear potential flow solver \cite{engsig2009efficient}. In this study, we use focused wave profiles with a Gaussian wave spectrum. We vary both the wave amplitude, the peak wavenumber and spectral bandwidth to cover a wide range of wave conditions (see details in Section \ref{sec:init}). We collect over 20,000 spatial-temporal wave evolution data points in this non-breaking wave database with full information on the wave fields at the surface including $\eta$, $u$, $w$, and all their first and second-order spatial derivatives, which are then used to formulate the final discovered equations.

\subsection{Re-Discovered non-breaking BC}
We trained our DISCOVER model with the non-breaking wave database. The proposed equation by symbolic regression model is:
\begin{equation}
\eta_{t,\text{Re-discovered}} =-1.043 \eta_x u+0.978 w,
\end{equation}
By comparing to the mathematically derived non-breaking BC in Equation \ref{eqn:fnpf}, the current framework identifies all the correct terms and the error of the coefficients for the leading term ($w$) is less than 2\%. 

\section{Wave Breaking Boundary Equation}\label{sec:ke}

We aim to discover an approximate surface elevation evolution equation which can be applied during wave breaking. Such an approach is clearly approximate and cannot capture every facet of wave breaking, such as droplet production from bursting bubbles, for example. However, we find that this simple assumption enables the development of a model which straightforwardly captures the key features of diverse breaking wave data sets. This also operates well with unseen breaking experiments and, as such, appears to capture the essential behaviour of wave breaking. 

\subsection{New Description of BC for Breaking Waves} \label{sec:raycast}
Due to the complex behaviour during wave breaking, it is challenging to define and obtain a robust description of the air-water boundary. As such, we start with a new description of the BC for a breaking wave by addressing two main characteristics: overturning and local air entrainment.

The overturning leads to multiple surface elevation points at a given horizontal position for Eulerian coordinates. To address this, we utilise a ray casting probing to interpolate the air-water boundary function, which mimics a dense gauge array during wave experiments and only samples the highest air-water boundary position for a pair of $(x,t)$. The new description of overturning waves provides unconditional stability for the surface elevation and all the spatial and temporal derivatives of the wave properties (see details in Section \ref{sec:ray}). 

The local air entrainment produces droplets and air cavities surrounded by water, which leads to multi-scale processes and also complex local profiles \cite{deike2015capillary}. We present a local perturbation approach to remove the localised small-scale processes, where we repeat the numerical wave experiments with perturbed initial conditions. Due to the chaotic nature of breaking waves \cite{wei2018chaos} at the small scale, statistical aggregation through these perturbed experiments can remove the effects of the local air entrainment. The full details of this perturbation approach can be found in Section \ref{sec:perturb}. 

\subsection{Breaking Wave Database} \label{sec:breaking_database}
We established a new breaking wave database by running high fidelity VoF method of wave groups with the Basilisk library \cite{popinet2009accurate} solving the two-phase 2D Navier Stokes Equation directly. We simulate a total of 45 breaking wave cases with 5 perturbations for each case. This provides over 300,000 spatial-temporal wave evolution data points being included in this breaking dataset within the breaking region. We collect the full information on the wave fields at the surface including surface elevation $\eta$, horizontal and vertical velocity $u$ and $w$, and all their first and second-order spatial derivatives. This information is then given to machine learning to assemble a new breaking boundary equation. Additionally, we have also included wave characteristic parameters -- peak angular frequency $\omega_p$, peak frequency $f_p$, and other relevant constants (gravitational acceleration $g$ and constants $\pi$). 

\subsection{Discovering new breaking BC}
We start with the classic and well-known kinematic boundary condition--fully nonlinear boundary condition (FNBC) for non-breaking waves:

\begin{equation}
    \eta_{t,\text{FNBC}} = - \eta_x u+  w,
    \label{eqn:fnpf}
\end{equation}
where $\eta$ is the surface elevation, subscript $_t$ and $_x$ denote the temporal and spatial derivatives, respectively, $u$ is the horizontal fluid velocity at the surface and $w$ is the vertical fluid velocity at the surface.

For breaking waves, however, once the free surface overturns it can no longer be represented by a single-valued function $\eta$ under the classic Eulerian representation. A Lagrangian representation will also break down when the overturning head reconnects to the free surface ahead. Nevertheless, many phase-resolved wave models use an effective vertical coordinate to represent the free surface in some sense \cite{engsig2009efficient}, even throughout the breaking process. Because such a representation in general does not match the true free surface, it is not guaranteed that the kinematic boundary condition needs to be met on $\eta$ during breaking, so Equation \ref{eqn:fnpf} cannot apply. Physically, the failure of Equation \ref{eqn:fnpf} corresponds to the allowance of a nonzero mass flux across the air-water interface, which is otherwise prohibited by the kinematic boundary condition; such a mass flux may take the form of air entrainment, for example. In any case, a correction to Equation \ref{eqn:fnpf} for use in such models is desirable. 

As such, we aim to discover a free-form PDE with a similar structure but specifically designed to capture wave evolution during breaking:
\begin{equation}
\eta_t=N\left(\eta, \eta_x, \eta_{x x}, u, u_x, w \cdots\right) \cdot \mathbb{\xi}, 
\end{equation}
where $N(\cdot)$ denotes a function term, which is related to fluid surface properties such as surface elevation $\eta$ and its spatial derivatives, velocities profiles and its spatial derivatives, $\xi$ denotes the coefficient vector, which is always sparse and has many zero elements in practice.

The open-form PDE discovery scheme, referred to as DISCOVER \cite{DISCOVER} is used to explore the possible forms of $N(\cdot)$ and coefficients for the unknown breaking wave BC, where the breaking wave database (Section \ref{sec:breaking_database}) provides the full training, validation and testing data. Only the spatio-temporal locations within the breaking region (see Section \ref{sec:class} for details) are used as training points, as the FNBC can model the rest of the non-breaking region. The details of the methods and hyperparameters are provided in the section  table \ref{tab:hyp_discover}. 

\begin{figure}
\centering
\includegraphics[width=1\linewidth]{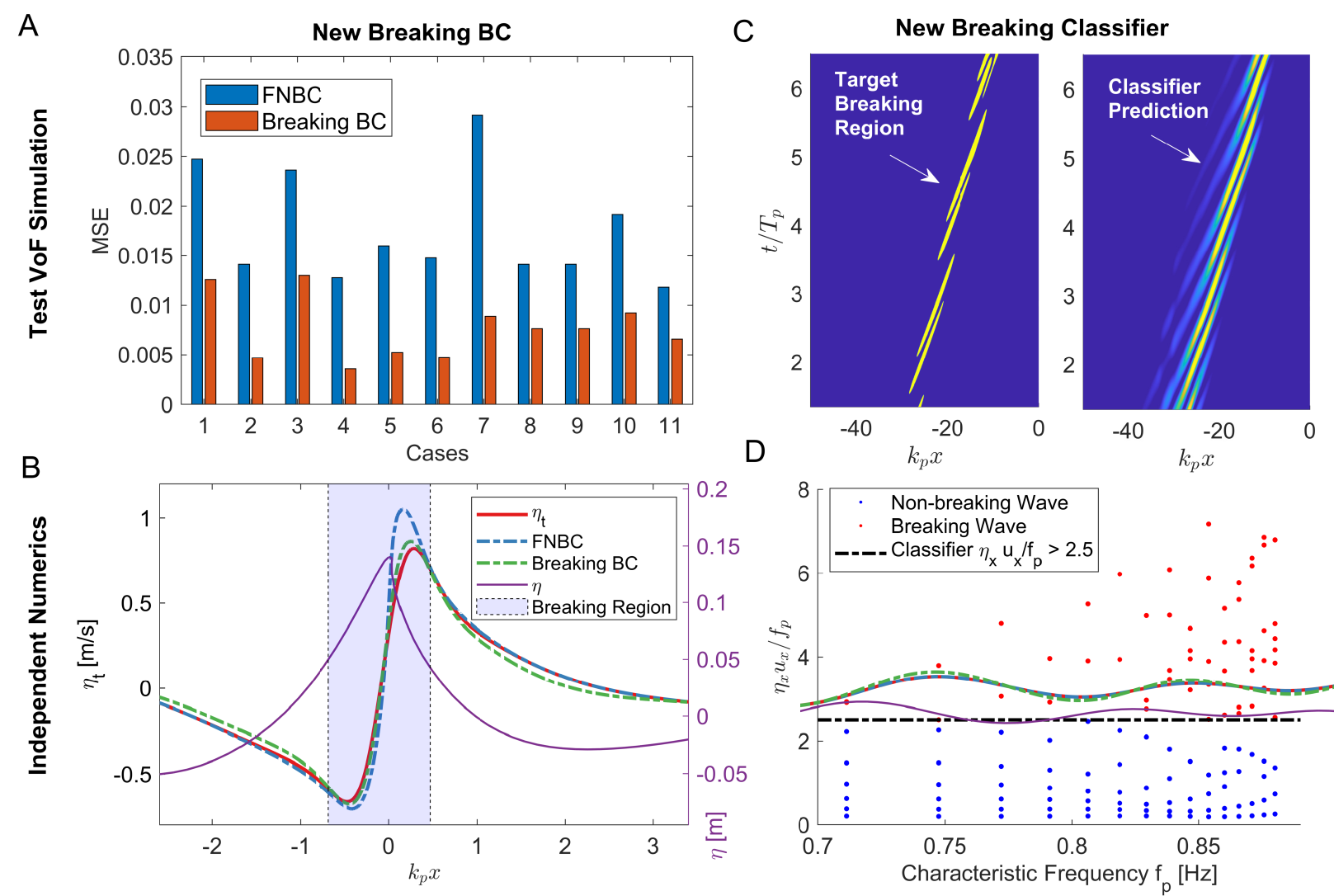}
\caption{\textbf{Accuracy and validation of new wave breaking boundary condition and breaking region classifier against test data and independent numerical simulation dataset} \textbf{Panel {A}} compares the existing fully nonlinear boundary condition (FNBC) and the new breaking boundary condition (BC) discovered for the test cases within the breaking region. 
\textbf{Panel {B}} compares the new breaking BC prediction during wave breaking against an independent unseen dataset published in \cite{mcallister2023influence}. The red line shows the target term $\eta_t$ in the spatial domain, the blue dashed line shows the calculation from the fully nonlinear BC (Equation \ref{eqn:fnpf}), the green dashed line shows the calculation from the new breaking BC (Equation \ref{eqn:breakingBC}) and the purple line indicates the surface elevation at this time instant. The shaded blue region represents the breaking region (Equation \ref{eqn:class}).
\textbf{Panel {C}} compares the new breaking classifier against an unseen test case. The breaking region is shown in yellow based on Equation \ref{eqn:region}, and the classifier shows the breaking region predicted through our new breaking classifier shown in Equation \ref{eqn:class}. 
\textbf{Panel {D}} applies the new breaking classifier to an independent unseen dataset published in \cite{mcallister2023influence}. the scatter with blue dots each represents the maximum classifier value achieved for a simulation without wave breaking, whilst the red dots indicate breaking event has occurred. The black dashed line shows the classification suggested by our model.}
\label{fig:mega}
\end{figure}

After running DISCOVER, the discovered new BC is
\begin{equation}
\eta_t=\underbrace{-\frac{g}{\omega_p} \eta_x}_\text{wave propagation term}+\underbrace{(-\frac{\omega_p}{\pi} B \eta_x)}_\text{dispersion correction term}+O(3),
\label{eqn:breakingBC}
\end{equation}
where $\omega_p$ is the peak frequency of the wave group, $\eta_x$ is the spatial derivative of the surface elevation $\eta$, and $B$ is the envelope of the surface elevation, which can be obtained as $B = \sqrt{\eta^2+\eta_H^2}$, so a fully nonlinear instantaneous wave envelope of the surface elevation, where $\eta_H$ is the Hilbert transform of the surface elevation.
\\
\\
\noindent \textbf{Accuracy and Validation}
We demonstrate the accuracy of this newly discovered breaking Equation in Figure \ref{fig:mega} panel \textbf{A}, where this new breaking BC achieved over 45 \% improvement in accuracy for all the test cases when compared to the existing non-breaking boundary condition.  Additionally, we apply this breaking boundary equation to an unseen independent numerical dataset published in \cite{mcallister2023influence} without reproduction, which is simulated with a Lagrangian fully nonlinear potential flow scheme \cite{dold1992efficient} in panel \textbf{B}. Within the breaking region before self-contact, our breaking BC predicts values which are much closer to the LHS of the Equation \ref{eqn:fnpf} (i.e. $\eta_t$) when compared to the fully nonlinear boundary condition. This shows strong evidence that the discovered equation is not a solution by coincidence, instead, this breaking BC can be applied directly to an unseen breaking dataset that is generated from a different numerical model with different initial conditions. Outside the breaking region, our discovered breaking BC still provides a reasonable prediction when compared to the FNBC, which can be traced back to the inherent mathematical structure of the equation (see further discussions in Section \ref{sec:math}). This also explains the excellent stability and mass conservation properties of this breaking BC shown in Section \ref{sec:stab}.

\section{Wave Breaking Classifier}\label{sec:class}
We now move on to the selection of the spatial and temporal locations between which we should apply the newly discovered new breaking BC. We define the target breaking region to be the locations where the FNBC  does not agree with the left-hand side of the equation:
\begin{equation}
\left| - \eta_x u+ w -\eta_{t}\right| > \chi \max(\left|\eta_{t}\right|),
\label{eqn:region}
\end{equation}
where the $\chi$ is the breaking region threshold. We use a value of 0.05 to mitigate the potential noise from the numerical differentiation. 

We modify the symbolic regression approach to perform classification by incorporating the L2 hinged loss function $L(a)$ as:
\begin{equation}
L(a)=\max \{0,1-a\}^2,
\end{equation}
where $a$ is the agreement value, which can be obtained as: $a = y\cdot\hat{y}$, where $y$ is the predicted label, and $\hat{y}$ is the target label. We detail the training dataset and hyperparameters in the supplementary materials. 

After running the modified symbolic classification, the discovered breaking classifier is:
\begin{equation}
\eta_x u_x/f_p > 2.5,
\label{eqn:class}
\end{equation}
where $f_p = \omega_p/(2\pi)$ is the peak frequency number. This threshold separates recurring wave breaking regions from non-breaking locations.
\\
\\
\noindent \textbf{Accuracy and Validation}
We demonstrate the accuracy of this breaking classifier in Figure \ref{fig:mega} panel \textbf{C}, where the new breaking classifier predicted region is compared to the actual breaking region (Equation \ref{eqn:region}) for an unseen test case obtained from VoF simulation. We report the similar spatial-temporal structure of this breaking region and consistent agreement throughout the test dataset.

The symbolic representation of this breaking region classifier allows us to simplify it into a single wave-breaking threshold similar to the existing dynamics and geometric breaking criteria \cite{saket2017threshold,barthelemy2018unified,derakhti2020unified,boettger2023energetic}, and we further test our breaking region classifier accuracy for this type of application using an independent dataset published in \cite{mcallister2023influence}. Figure \ref{fig:mega} panel \textbf{D} shows the maximum classifier value $u_x\eta_x$ of 149 wave groups in this dataset with varying steepness, bandwidth and characteristic frequencies (see Section \ref{sec:mark_data} for details). We report a separation between the breaking and non-breaking wave groups by the wave-breaking threshold simplified from the discovered breaking region classifier, which indicates this classifier can make accurate predictions out of the box for unseen datasets. 
\\
\\
\noindent \textbf{Comparison to theory}
The simple symbolic representation of the wave-breaking threshold implied by the breaking region classifier also allows us to convert the expression into a similar format of the classical limiting waveform of Stokes' theory \cite{stokes1880considerations}:
\begin{equation}
\left|\min \left(\eta_x\right)\right| > 1 / \tan (\pi / 3)(\approx 0.58).
\end{equation} 
To achieve this, we approximate $u_x$ following the linear theory as $u_x \approx 2\pi f_p \eta_x$. This allows us to further simply our breaking classifier as:
\begin{equation}
\max \left( u_x \eta_x/f_p \right) > 2.5  \Rightarrow \max \left(\eta_x^2 \right) > 2.5/(2\pi).
\end{equation}
Since the $\eta_x$ is negative during the breaking initiation stage, we have:
\begin{equation}
\left|\min \left(\eta_x\right)\right| > \sqrt{2.5/(2\pi)}(\approx 0.63), \label{eqn:stokes_alike}
\end{equation}
which provides a threshold value similar to the limiting waveform of the existing Stokes' theory. We note that this conversion involves estimating velocity fields from surface elevation with linear theory ($u_x \approx 2\pi f_p \eta_x$), which inevitably introduces extra errors for such a non-linear process, and likely is the reason why this threshold after the conversion is slightly larger than the Stokes' theory limit. This conversion, on the other hand,  highlights the advantage of such a symbolic approach -- allowing full flexibility on modification and re-formulation of the results, and easy integration with existing knowledge and theories using mathematical tools.

\section{New Wave Breaking Model}\label{sec:sim}
Based on the new breaking boundary equation and the new breaking classifier, a time-marching simulation scheme can be developed to approximate the evolution of the breaking wave free-surface. 

One key limitation of this model is that the current breaking BC only approximates surface elevation, whereas the breaking classifier requires information from velocity (i.e. $u_x$). In this study, the required $u_x$ term is taken directly from the numerical simulation results so assumed to be known prior. However, this term is only used to identify the breaking region (i.e. where to apply the new breaking BC). The modelling of the breaking surface elevation evolution is done through new breaking BC, which does not require any prior information. Further analysis has shown that the exact boundaries of this breaking region have a minimal effect on the simulated surface elevation profile. 

The new breaking model uses a unified wave-breaking boundary equation, which computes the non-breaking evolution from the fully nonlinear BC and applies the new breaking BC within the breaking region as predicted by the breaking classifier:
\begin{equation}
\eta_t= \underbrace{\left[- \eta_x u+ w\right]}_{\textrm{FNBC}} (1-\varepsilon) + \underbrace{\left[-\frac{g}{\omega_p} \eta_x -\frac{\omega_p}{\pi} B \eta_x\right]}_{\textrm{New Breaking BC}} \varepsilon, 
\label{eqn:full_ke}
\end{equation}
where $\varepsilon$ is the evolutionary breaking region classifier, which also takes account of wave propergation and dispersion effects from previous time steps:
\begin{equation}
\varepsilon = 1 - \prod_{ l = 0}^{l = m} \prod_{ j = 1 }^{j = n}\left[1-\mathbb{R}(u_x(t_l,x_j)\eta_x(t_l,x_j) /f_p - 2.5)\right], \label{eqn:evo_b}
\end{equation}
where $\mathbb{R}(\cdot)$ denotes an activation function, which smoothly connects breaking and non-breaking regions, a moving average window is used herein. The value $m = 3$ is the number of previous time steps to consider for change in breaking region due to dispersion. The value $n$ is the number of frequency components considered (in this study, $n$ is chosen to cover 95\% of the potential energy of the wave) and $x_j$ provides the new spatial position of breaking region at $l^{\textrm{th}}$ of the previous time step. This can be obtained through linear evolution of the $j^{\text{th}}$ wave component with frequency of $\omega_j$:
\begin{equation}
x_j = (x + g/\omega_j l\Delta t),
\label{eqn:evo_x_po}
\end{equation}
where $\Delta t$ is the time step of the simulation (the detailed illustration of this evolutionary breaking region is shown in Section \ref{sec:prop_breaking_region}).

\begin{figure}
\centering
\includegraphics[width=0.85\linewidth]{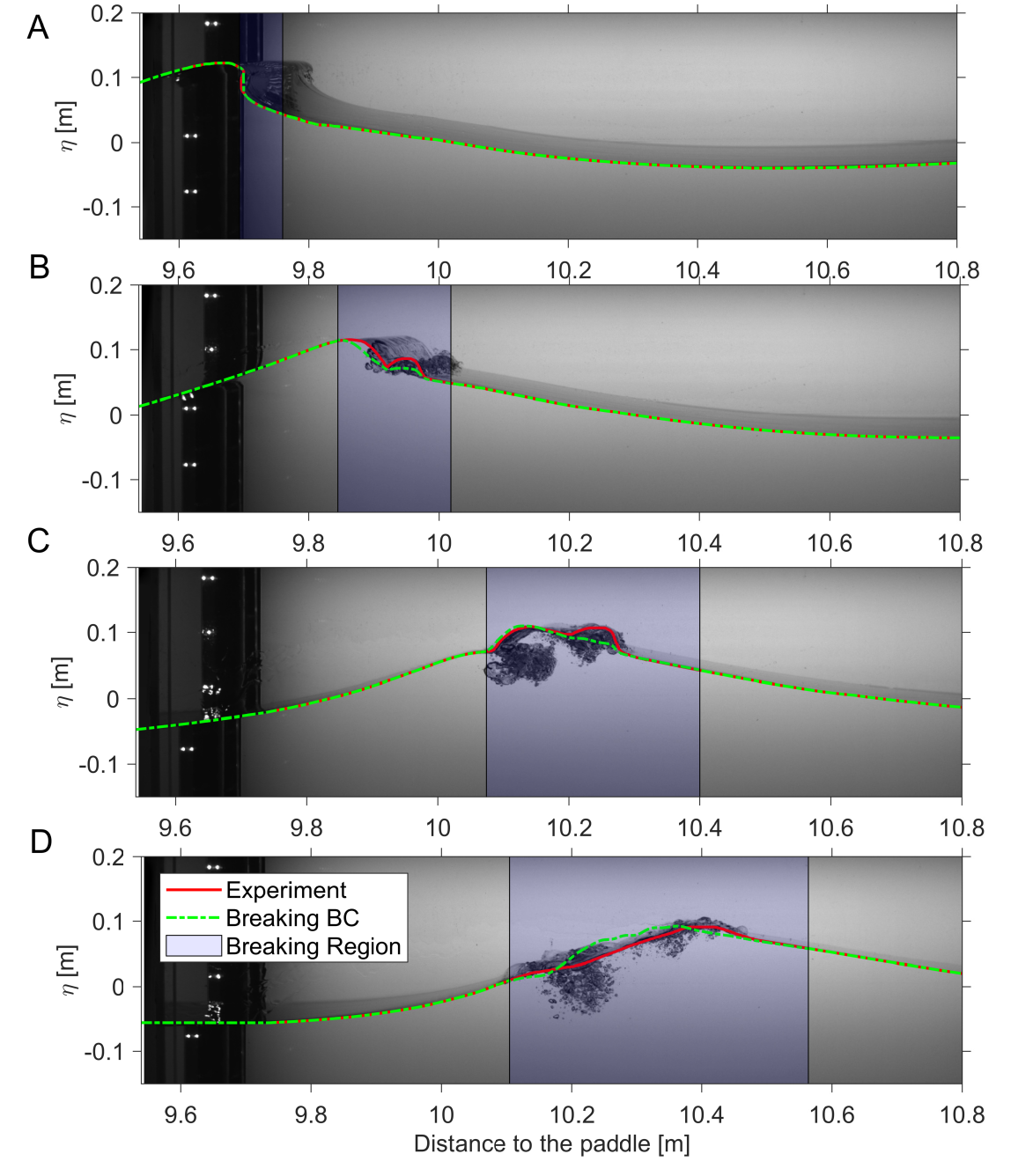}
\caption{\textbf{Validation of new breaking wave model against unseen wave experiment.} We first identify the breaking region according to the wave-breaking classifier in Equation \ref{eqn:evo_b}. The new breaking BC is then applied within the breaking region to model the surface elevation evolution during breaking. A classical 4th order Runge–Kutta method is used to evolve the new breaking BC in time with a time step of 1/320 $T_p$. The wave evolution in the rest of the non-breaking region is adopted from the experimental results to avoid edge effects. Panels \textbf{A} - \textbf{D} compare our breaking BC model with experimental results at time instances of 0.00 seconds, 0.15 seconds, 0.35 seconds and 0.50 seconds, respectively. }
\label{fig:exp}
\end{figure}

A classical 4th order Runge–Kutta method (RK-4) \cite{butcher1976implementation} is used to evolve the Equation \ref{eqn:full_ke} in time with a time step of $1/320 T_p$, and rigorous convergence tests are performed to ensure numerical convergence. This is essentially very similar to well-known simulations of the KdV \cite{dutykh2014numerical}, and Burgers equation \cite{mittal2012numerical}, and the velocity profiles used in FNBC are directly obtained from VoF simulations. 
\\
\\
\noindent \textbf{Validation against physical experiment} We test our new breaking wave model by applying it to independent physical experimental results as shown in Figure \ref{fig:exp}. The experimental data were obtained in a wave-flume in Imperial College London and specific details are shown in Section \ref{sec:ic_exp}. 

We first use an automated wave profile detection algorithm to detect the surface wave profile (detailed in Section \ref{sec:extract}). We identify the wave-breaking region following the Breaking Classifier in Equation \ref{eqn:evo_b}, where the horizontal velocity term is approximated linearly as $u_x \approx 2\pi f_p \eta_x$. The wave evolution in the rest of the non-breaking region is adopted from the experimental results to provide surface elevation information outside the breaking region and avoid edge effects. 

We directly compare our breaking BC model against the experimental results in Figure \ref{fig:exp}, where the new breaking BC is applied within the blue-shaded breaking regions. We report excellent agreement at the initial stage of the wave breaking. During the splashing stage, the new breaking model accurately captures the overall movement of the bulk water, consistent with physical experimental results. The small-scale splashing behaviour, however, is not captured by our model. This is expected as our breaking model focused on the evolution of the bulk water motion during wave breaking at the wavelength scale, whereas these droplets and local cavities at small scales are largely mitigated by perturbation aggregation in the data (see details in Section \ref{sec:raycast}).

\section{Mathematical Interpretation and Physical insights} \label{sec:math} 
\subsection{Mathematical Interpretation}
The knowledge discovery scheme is intended to provide full mathematical interpretability of the output. The leading wave propagation term in the new breaking BC (Equation \ref{eqn:breakingBC}) can be shown as a narrow-band approximation of the original leading $w$ term in the FNBC:
\begin{equation}
    \eta_t \approx -{g}/{\omega_p} {\eta_x}.
\end{equation}
This connection highlights the mathematical consistency of this BC and also allows further modification to include finite water depth effects (e.g. $w \approx -{g}/{\omega_p} \tanh(k_p d) {\eta_x}$), where $k_p$ is the peak wavenumber and $d$ is the water depth. 

The second dispersion correction term in new breaking BC (Equation \ref{eqn:breakingBC}) provides a nonlinear correction to the dispersion relationship, which makes waves with larger amplitude travel faster than the smaller waves of the same frequency and is of similar behaviour to the nonlinear dispersion relationship derived in the classical theory for non-breaking waves \cite{stokes1880considerations}.  However, the amount of correction ($\omega_p/{\pi} B$) in the phase speed is proportional to the envelope $B$, which is different to the non-breaking wave theories at $O(A^2)$, where $A$ is the wave amplitude. This is likely to be attributed to the slowing behaviour of a crest prior to breaking, which has been previously reported in numerous studies \cite{banner2007wave,grue2012orbital,khait2018kinematic}. 

This new breaking BC also suppresses the dispersion when compared to the non-breaking wave evolution, where the first propagation term is non-dispersive by its nature and the second term is only amplitude dispersive but not in frequency. This suppression of dispersion automatically leads to a shock-type wave propagation for the wavefront, where previous studies show a similar treatment of wave breaking without any ad-hoc parameterisation in shallow water \cite{TISSIER201254,ORSZAGHOVA2012328,SHI201236,BONNETON20071459}.

\subsection{Physical Insights} \label{sec:phy}
The full mathematical interpretability of the output provided through the knowledge discovery scheme also suggests new physical insights of the breaking wave. For the non-breaking evolution, the current discovery framework shows the capability of re-discovering the fully nonlinear BC (see details in Section \ref{sec:redis}). However, for breaking evolution, a new form that only includes surface elevation properties has been found. The difference between non-breaking and breaking formulations will provide insights into breaking events.

\begin{figure}[t]
\centering
\includegraphics[width=1\linewidth]{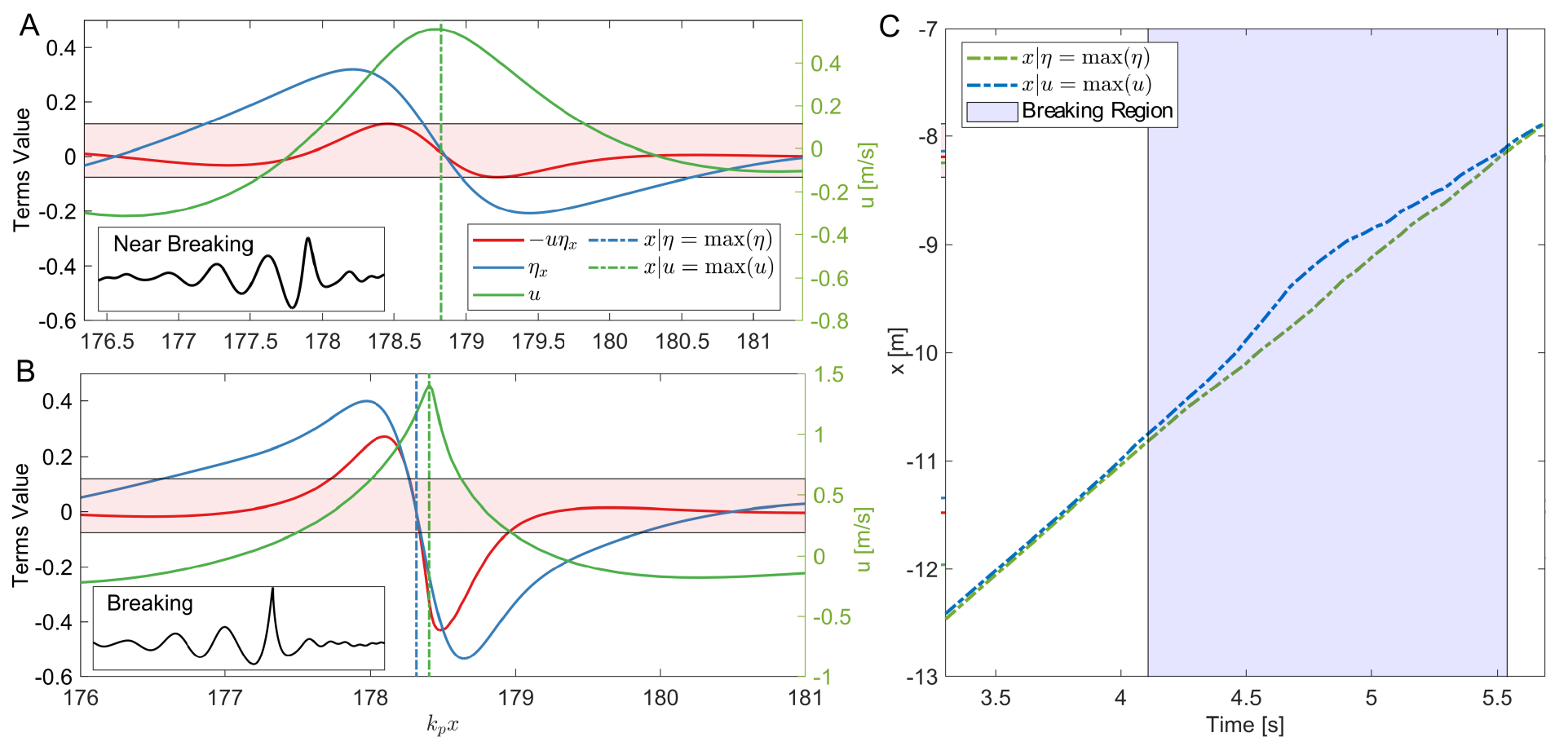}
\caption{\textbf{The new breaking boundary equation suggests a new phenomenon during wave breaking} --- the misalignment between the maximum position of surface elevation $\eta$ shown in blue and here $u$ is the horizontal velocity on the free-surface shown in green. This can be demonstrated by comparing a near-breaking wave group in \textbf{A} and a breaking wave group in \textbf{B}. The misalignment only occurs in breaking cases where the maximum location of $\eta$ (i.e. zero crossing point spatial derivative $\eta_x$) separates from the maximum location of horizontal velocity $u$. This leads to an unseen increase in the amplitude of the $\eta_x u$ term shown in the red line with non-breaking amplitude in the red shade. This misalignment can be further demonstrated in \textbf{C}, which only occurs during wave breaking in space and time. Wave data for panel \textbf{A} and \textbf{B} is obtained directly from independent numerical dataset \cite{mcallister2023influence}, and panel \textbf{C} is from high fidelity VoF method.}
\label{fig:physics}
\end{figure}

To explore the underlying mechanics, we explore both the test dataset from VoF simulation and the independent validation dataset previously published in \cite{mcallister2023influence} generated from a different numerical scheme. We first investigate the validity of the fully nonlinear BC for near breaking and breaking cases in Figure \ref{fig:physics} with data from \cite{mcallister2023influence}. For the near breaking case, as shown in panel \textbf{A}, surface elevation $\eta$ is closely aligned with the horizontal velocity $u$, leading to the overlapping of maximum positions for both $\eta$ and $u$. This alignment between $\eta$ and $u$ limits the amplitude of the $- u\eta_x$ term in the FNBC, where the zero value of $\eta_x$ corresponds to the maximum value of $u$. However, for a breaking case that is close to overturning shown in panel \textbf{B}, the misalignment between the $\eta$ and $u$ leads to significant growth in the $- u\eta_x$ term, which causes an increased value of $\eta_t$ around the maximum surface elevation position. This indicates a more pronounced change of $\eta$ around the crest and can be closely linked back to the breaking behaviour during the inception stage \cite{mcallister2022wave}. We further confirm this misalignment in the spatial-temporal domain in panel \textbf{C} from the VoF simulation with a different breaking wave case, where significant decoupling effects can be also observed throughout the breaking evolution. This is superficially similar to the acceleration mismatch previously reported in \cite{pizzo2017surfing}. As such, we believe this decoupling between surface elevation and velocity suggested by new breaking BC is a robust characteristic for breaking waves within the proposed effective air-water boundary in Section \ref{sec:raycast}. 

The new breaking BC avoids this misalignment by using an equivalent approximated term $-{g}/{\omega_p} {\eta_x}$ to replace the original $w$ term in the leading order and avoids the use of $u$ terms to perform the dispersion correction in the second term.

\section*{Limitations} We highlight the following limitations: 
\begin{itemize}

    \item The ray-casting description of the boundary and the perturbation aggregation effectively removes the fine details of the breaking structure (i.e. the splashes and the cavities) after the breaking inception. This leads to the newly discovered BC focusing on the kinematics of the majority of the water, whereas the BC is expected to be very different for mixing behaviour between air and water on a smaller scale (for example, the formation of jets \cite{mcallister2022wave} and air entrainment \cite{iafrati2014modeling}).

    \item Most of the training cases in this study are in deep water and with spilling breaking type. This would suggest the current breaking model cannot be directly used for cases where breaking is governed by other physics, such as depth-induced breaking in coastal regions \cite{smit2013depth}. The overturning-impacting stages are also much more distinguished for these waves and likely to require a separate treatment for each stage.
    
    \item We only have uni-directional wave group simulations hence the discovered BC is only limited to a 2D breaking model. While we expect extra terms to be involved for directional spread waves, the breaking mechanism we discovered for unidirectional waves should still provide valuable insights into breaking waves in 3D provided the directional distribution of energy is not so large that other physics will become important \cite{mcallister2024three}. %This leads to the design-critical extreme waves closer to the unidirectional limit. 

    \item The training dataset in this study is generated from numerical simulations, which will not exactly replicate real wave behaviour. To mitigate this potential limitation, we have performed rigorous convergence tests for current simulations, and a physical wave experiment to validate the new breaking wave model.
    
\end{itemize}

These limitations could potentially limit the direct application range of this new breaking wave BC. The validation results, however, suggest the discovered equation still shows some predictive ability for unseen datasets, including for unseen 2-D experimental datasets. Hence, we are confident our approach has general applicability. 

\section*{Next Steps}

The work presented here is intended to open a new approach to modelling and revealing physical insights of breaking waves. It is not an end in itself but we hope to give others a new perspective to explore challenges in wave breaking research. Some next steps for the work include the following.
\\
\\
\noindent \textbf{Breaking Strength Indicator} The breaking strength indicator describes the intensity of the breaking wave \cite{banner2007wave}, which is often used to categorize breaking behaviour and to predict breaking inception \cite{perlin2013breaking}. Various breaking intensity parameters have been proposed and compared against experiments \cite{duncan1994formation} and numerical results \cite{cui2022large}. These parameters, however, require the full wave kinematics profiles beneath the water surface, which presents a significant computational task for engineering applications. With the new BC and symbolic classification tools proposed in this study, we believe one future work direction is to distil a new breaking strength indicator using the information at the free surface only. This will improve the impact force predictions from these breaking waves on offshore infrastructure structures and renewable energy devices \cite{wienke2005breaking}.
 \\
\\
\noindent \textbf{Directionally Spread Waves}
Real ocean waves, both breaking and non-breaking, are directionally spread \cite{adcock2009estimating}, occur in the presence of surface currents and direct wind forcing \cite{fritts1994gravity}. Therefore, we expect the breaking behaviour to be more complicated in the open ocean. Indeed, numerical VoF simulations in 3-D lead to a dramatic increase in computational cost for training dataset generation, and the 3-D BC is expected to be more complicated and likely to include partial derivatives in the transverse direction. All the critical steps in our current study are not constrained to unidirectional waves and we therefore believe that our approach can be extended to spreading waves. The current boundary equation provides a starting point for this work.
\\
\\
\noindent \textbf{Shallow Water Breakers}
Breaking waves also play an important role in coastal protection and surfing activities \cite{salmon2015modeling}. These breaking waves, however, are usually initiated through a slightly different mechanism and also have a different breaking type \cite{smit2013depth}. Our approach can potentially provide a new opportunity to explore these mechanisms behind shallow water breakers and reveal new physical insights. The current boundary equation for breaking waves in deep water also provides a great starting point.

\section{Methods}\label{sec11}

\subsection{Symbolic Regression}\label{sec:sr}
We employed an advanced symbolic regression method called DISCOVER to directly extract governing equations from data, resulting in explicit mathematical expressions \cite{DISCOVER}. This method is, fundamentally, an interpretable machine-learning approach that transforms the mapping relationships between various physical quantities into equation-based surrogate models. It combines the high accuracy of surrogate models with the interpretability of equation expressions, facilitating theoretical research for scientists.

The core challenge of knowledge discovery lies in the computer’s inherent difficulty in understanding and processing mathematical equations directly, which hampers the rapid identification of underlying governing equations—essentially the problems of equation representation and equation optimization. DISCOVER addresses these challenges by first using symbolic mathematics to establish a one-to-one mapping between any form of partial differential equation and a binary tree, converting complex mathematical equations into a format that computers can efficiently handle, thus solving the issue of equation representation. Then, using reinforcement learning, an LSTM agent with structural awareness iteratively generates the binary tree that best describes the input data, which is subsequently converted back into an equation. The core idea of DISCOVER is to bridge the gap between equations and data through symbolic mathematics and binary trees, leveraging reinforcement learning to efficiently uncover the governing equations underlying the data.

DISCOVER has demonstrated the ability to identify various complex structured equations, such as the differentiation of composite functions or fractional structural equations, making it well-suited for this research. Additionally, unlike traditional symbolic regression methods, DISCOVER does not rely on a predefined closed candidate set, offering greater flexibility and the ability to generate previously non-existent equation terms, making it more effective in discovering novel governing equations in unknown processes.

In this study, we did not force the unit to be consistent during the symbolic regression operation. Instead, we non-dimensionalised the discovered terms with appropriate characteristic constants i.e. peak wavenumber, peak frequency, and re-perform the final sparse regression to determine the final coefficients. 

In addition to the DISCOVER package, we have also used the PySR \cite{cranmer2020discovering,cranmer2023interpretablemachinelearningscience} as an alternate package to perform symbolic regression classification for the breaking region identification problem. PySR is a symbolic regression package based on genetic programming \cite{grefenstette1993genetic}, which selects the best candidate out of an ensemble before mutating and recombining them into the next generation. The PySR package represents the mathematical expressions of a tree of constants and elementary symbols, allowing the package to discover expressions of unbounded complexity. PySR utilises a similar centric metric as DISCOVER to evaluate the fitness of an equation to the dataset based on parsimony, where the improvement in the predictive performance is evaluated against the model complexity. This provides the maximal information gain from each of the increases in the model complexity and prevents model over-fitting by penalising the tiny improvement gain with over-complex equations. All the results in the final paper for the new breaking BC were derived using DISCOVER as for this problem it had better efficiency. The breaking classifier is achieved with PySR due to the DISCOVER is developed and optimised mainly for regression problems.

\subsection{Numerical Experiments}\label{sec:numeri}
\subsubsection{Basilisk solver}
In this study, we use the Basilisk library to simulate two two-dimensional breaking wave evolution, which solves the two-phase incompressible Navier–Stokes equations directly with surface tension following \cite{farsoiya2022direct,liu2023wave,mostert2020inertial}. The Basilisk, and its predecessor (Gerris), flow solver \cite{popinet2003gerris,popinet2009accurate} have been used to simulate various partial differential equation systems, including breaking waves \cite{liu2023wave,mostert2020inertial}. The governing equations can be written as:
\begin{equation}
\begin{gathered}
\frac{\partial \rho}{\partial t}+\nabla \cdot(\rho \boldsymbol{u})=0 \\
\rho\left(\frac{\partial \boldsymbol{u}}{\partial t}+\boldsymbol{u} \cdot \nabla \boldsymbol{u}\right)=-\nabla p+\nabla \cdot(2 \mu \boldsymbol{D})+\rho \boldsymbol{g}+\sigma \kappa \delta_s \boldsymbol{n} \\
\nabla \cdot \boldsymbol{u}=0,
\end{gathered}
\end{equation}
where $\rho, \boldsymbol{u}, \mu, \sigma, \boldsymbol{D}, \boldsymbol{g}$ are the water density, velocity vector, dynamic viscosity, surface tension, deformation tensor and gravitational acceleration vector. A fraction field $\mathbb{T}(\vec{x}, t)$ is used to vary the density and viscosity at the interface between air and water, which is zero in the gas phase and becomes unity in the liquid phase. The $\delta_s$ is a Dirac delta that embedding surface tension effects into the water-air interface, and $\kappa$ is the interface curvature, and $\boldsymbol{n}$ is its unit normal vector of the air-water interface.
\subsubsection{Numerical Experiment Setup}
We consider breaking focused wave groups in deep water. The relevant parameters for the simulation are the
liquid and gas density $\rho_w$, $\rho_a$, dynamic viscosities of liquid and gas $\mu_w$, $\mu_a$, the surface tension $\sigma$, and gravitational acceleration $g$. We set water depth $h_0$ to be 5 meters in this study, which is sufficiently deep ($h_0kp \approx 8$) so it does not affect the wave evolution during breaking. The values of numerical experiment setup parameters are detailed in Table \ref{tab:param}.

Our numerical simulation mimics a physical wave tank with a length $l$ of 66 meters, which is large enough to prevent waves from reaching the boundary, thus eliminating reflections from the boundary. The Reynolds Number and Bond number also represent an experimental scale at which the breaking evolution is independent of both of these non-dimensional parameters \cite{IAFRATI_2009}. We define the numerical resolution as the ratio between the smallest cell size in the simulation and the characteristic wavelength following \cite{mostert2020inertial}, $\Delta=l / 2^L$, where $L = 16$ is the maximum level of refinement. We also enforce the maximum level of refinement at the free surface throughout the simulation to ensure a better representation of the interface.

An adaptive mesh refinement scheme is used to allow efficient allocation of computational resources to the active portions of flow. A detailed description of the scheme can be found in \cite{van2018towards}. The refinement scheme adapts the 2D quadtree mesh based on the velocity and VoF tracer fields. We also enforced the maximum level of refinement at the air-water interface throughout the simulation. Rigorous convergence tests have been performed to ensure that all the fluid quantities at the surface that are included in the breaking database  (e.g. velocities and their time and spatial derivatives) and the surface elevation itself are not changing over 1\% when further increasing the numerical resolutions or the Courant–Friedrichs–Lewy number (see Figure \ref{conv} for details).

\subsubsection{Wave Initialisation}\label{sec:init}
The wave initialisation can be separated into two parts: fully non-linear simulation pre-conditioning and VoF simulation with Basilisk. This fully nonlinear pre-conditioning avoids the error wave propagation within the numerical tank \cite{schaffer1996second}, which allows the VoF simulation to be started at a time instance that is closer to breaking.

We use a fully non-linear code described in detail in \cite{engsig2009efficient} to obtain the velocity potential and the surface elevation closer to break, where a unidirectional wave profile with a Gaussian wave spectrum following \cite{gibbs2005formation} is specified as:
\begin{equation}
    S(k) = \exp \left(\frac{-\left(k-k_{\mathrm{p}}\right)^2}{2 k_{\mathrm{w}}^2}\right),
\end{equation}
where $S(k)$ is a Gaussian wave spectrum as a function of wavenumber $k$, $k_{{p}}$ is the peak wavenumber and $k_{\mathrm{w}}$ is the spectral bandwidth. We compute the surface elevation $\eta(x, t)$ and velocity potential $\varphi(x, z, t)$ from the wave spectrum as:
\begin{equation}
\begin{aligned}
& \eta(x, t)=\sum_{n=1}^N a_n \cos \left[k_n\left(x-x_c\right)-\omega_n\left(t-t_c\right)\right] \\
& \varphi(x, z, t)=\sum_{n=1}^N a_n \frac{\omega_n}{k_n} \frac{\cosh k_n(z+h)}{\sinh k_n h} \sin \left[k_n\left(x\right)-\omega_n\left(t-t_c\right)\right]
\end{aligned}
\end{equation}
where $\eta$ is the surface elevation, $\phi$ is the velocity potential, $a_n$ is the amplitude of $n-th$ component $\omega_n$ is the angular frequency, $k_n$ is the wavenumber, $t_c = -20 T_p$ is the focused time, $z$ is the vertical coordinate, and $h$ is the water depth. In this study, focused wave groups are generated with the peak period range $1<T_p<2$ seconds, $0.0046<k_{\mathrm{w}}<0.00575$ m$^{-1}$, and $0.17< Ak_p<0.25$, where $T_p$ is the peak period and $Ak_p$ is the non dimensionalised wave amplitude at linear focus. A total of 225 simulations (including perturbed ones) are generated for the breaking database, which covers a wide range of various breaking wave group profiles.

We specify the surface elevation and velocity potential spatially for the fully non-linear simulation, where the exact second-order corrections following \cite{dalzell_note_1999} are used in this study to mitigate the error wave contamination. When the wave is close to breaking, we stop the fully nonlinear simulation pre-conditioning scheme and use the simulated results as the final initial condition for the VoF method. To achieve this, we specify the tracer function and the velocity fields of the water in the 2D quadtree mesh using a Fourier-based data assimilation approach. Over 20000 frequency bins are used to ensure the assimilation accuracy. This allows us to initialise VoF directly from the fully nonlinear simulation to mitigate further error wave contamination. The total simulation time in Basilisk is fixed at 10 seconds, which allows around 5-6 periods of breaking evolution.

\subsection{Ray casting description of air-water interface}\label{sec:ray}
We need a new description of Boundary Condition (BC) for overturning waves that avoids non-single values of surface elevation in an Eulerian specification. We use a ray casting description of the air-water boundary in two steps: reconstructing the multi-fluid interface and ray casting probing.

We first reconstruct the interface between the air and water through a volume of fluid (VoF) method, where a tracer function $\mathbb{T}(\vec{x}, t)$ is defined through the volume fraction of a given fluid in each computation cell, where $\vec{x} = (x,z)$ is the position vector. The density $\rho$ and viscosity $\mu$ can hence be obtained as $\rho(\mathbb{T})=\mathbb{T} \rho^w+(1-\mathbb{T}) \rho^a$ and $\mu(\mathbb{T})=\mathbb{T} \mu^w+(1-\mathbb{T}) \mu^a$, where $^a$ and $^w$ denotes the property of the two fluids (water and air) respectively. A detailed description of the numerical methods and surface reconstruction can be found in \cite{popinet2009accurate}. 

We further utilise the ray casting probing to interpolate the tracer function $\mathbb{T}(\vec{x}, t)$ as:
\begin{equation}
\eta (x_0,t) = p - d - \frac{\vec{n}\cdot(\vec{x}_{\textrm{ori}} - \mathbb{T}(\vec{x}_{\textrm{int}}, t))}{\vec{n}\cdot\vec{r}_{\textrm{dir}} },
\end{equation}
where $\eta(x_0,t) $ is the interpolated surface elevation at $x = x_0$, and $\vec{x}_{\textrm{ori}} = \sqrt{x_0^2+p^2}(x_0,p)$ is the position vector of casting origin and $p$ is the vertical coordinate of casting origin, $d$ is the still water level, $\vec{r}_{\textrm{dir}} = (0,-1)$ is the ray-casting direction vector, $\vec{n}$ is the normal vector of the local tracer function $\mathbb{T}(\vec{x}_{\textrm{int}}, t)$, and $\vec{x}_{\textrm{int}} = (x_0,z_0)$ denotes the interpolation point, where $z_0 = {\text{argmin}}( j \cdot (\mathbb{T}((x_0,z), t)$, and $j$ is the direction vector in $z$ direction. This sampling approach mimics the wave probes used during the wave experiment and samples the upper layer of the water-air interface as the surface elevation value.

We use a total of 16,384 probes across the numerical wave flume, where rigorous grid convergence tests have been performed to ensure no clear further improvement can be observed for further increasing the number of probes.

\subsection{Local perturbation to remove small scale effects}\label{sec:perturb}

In this section, we detail the local perturbation approach we used in the new description of Boundary Condition, where the aim is to remove the small-scale effects such as the splashing and cavities at the air-water interface. These effects naturally arise from the nature of breaking waves \cite{wei2018chaos}, and introduce significant challenges in defining a stable and continuous Boundary Condition to describe the bulk water movement. 

This is done through a small perturbation on the wave initial conditions specified in Section \ref{sec:numeri}\ref{sec:init}. For example, the $j^{\textrm{th}} $perturbed quantity ${q}^{ \pm \delta j} $ can be obtained as: 
\begin{equation}
{q}^{ \pm j \delta }=q_1 \pm \delta j q_1+q_{22} \pm 2 \delta j q_{22} \cdots, \label{eqn:purt}
\end{equation}
where $\delta = 0.005$ is the small parameter, which controls percentage variation of the quantity, $q_1$ is the linear part of the quantity and $q_{22}$ is the second order part, which can be obtained following \cite{dalzell_note_1999}. Following Equation \ref{eqn:purt}, we perturb both the surface elevation and velocity potential to avoid mismatch at the boundary. The $j$ controls the number of initial conditions generated per case during the perturbation process. In this study, we limited $j = 2$ due to computational resources, which gives a total of $2j+1$ initial conditions per case. We then run a full VoF scheme on each of these perturbed initial conditions, which gives calculated quantity $q_c^{ \pm j \delta }$ corresponding to the $j^{\textrm{th}} $perturbed initial condition. Finally, we run a median operation to remove the small-scale chaotic behaviour during the wave breaking across all the perturbed and unperturbed initial conditions:
\begin{equation}
q_c =\operatorname{Med}\left[q_c^{(0)} + q_c^{ \pm  \delta} +q_c^{ \pm 2 \delta }+ \cdots +q_c^{ \pm j \delta }\right] \\ \text { for } j=1,2,3,  \cdots
\end{equation}
where the $q^{(0)}(x,t) $ is the undisturbed quantity.

\subsection{Breaking wave region evolution due to wave propagation}\label{sec:prop_breaking_region}

In this study, we constrain the application of the new breaking BC to be only within the breaking region. To complete a breaking wave group simulation, the non-breaking BC (i.e. FNBC) is needed to be applied at regions that never being affected by the breaking. As such, we need to consider both the breaking region identified by the classifier at the current time step and also possible contamination propagated from the previous time steps due to breaking. 

Here, we illustrate how Equations 11 and 12 in the original manuscript work to update the wave-breaking region due to the contamination propagated from the previous time steps. We now consider a breaking grid point happens at $l$ time steps before the current time step in Figure \ref{breaking_region}. Due to the wave breaks at the grid point as identified by the classifier, we shall consider some regions of the current time step ($x_1$ to $x_n$) contaminated due to the previous breaking. This also ensures that the breaking and non-breaking BCs are not switched very frequently, which leads to significant numerical instabilities. We decide the region at the current time step contaminated based on the linear theory, where $l\Delta t$ is the time difference between two time instances. For a wave with a frequency of $\omega _i$, the traveling distance between $l\Delta t$ will be $g/\omega _i l\Delta t$ with the speed of $v = g/\omega _i$. We consider the frequency range covering 95\% of the energy of the wave group in this study. This evolutional break region due to contamination is then combined with the breaking region classifier at the current time step in Equation 11. The final breaking region is defined as either the contamination region affected by the wave breaking in previous time steps or the new breaking region identified from the classifier.

\subsection{Numerical stability and conservation of mass for the breaking BC}\label{sec:stab}
We examine the numerical stability of the new breaking BC when used as the only governing equation in an RK4 time-stepping scheme. We simulate the spatial-temporal evolution of a non-breaking wave group in Figure \ref{fig:regular}. We observe stable numerical results throughout the simulation for over 5 peak periods, whereas a typical wave-breaking event only lasts for less than 2 peak periods. We also report almost perfect mass conservation properties of the discovered new breaking BC, where the variation in the total mass of the water throughout the entire simulation is less than 0.1\%.

\subsection{First propagation term under linear theory}
We now derive the first propagation term in the new breaking BC from linear wave theory. We follow the Stokes type of expansion with a narrow-banded approximation and expand the nonlinear elevation with higher-order harmonics as:
\begin{align}
    \eta &= \eta_1+\eta_2+\eta_3+\cdots
    \label{eqn:eta_expand}
\end{align}
where $\eta_n$ is the nth order of harmonics, and the higher-order terms can then be obtained as depth-dependent functions of the linear terms as
\begin{align}
\eta_2 &=S_{22} \ \ k_p(\eta_1^2 - \eta_{1H}^2),
\label{eqn:s22}
\end{align}
where subscript $H$ denotes the Hilbert transform    \citep{birkhoff1984establishment}, which phase shifts all the Fourier components of a signal by 90$^\circ$, $k_p$ is the peak wavenumber, and $S_{22} $ is the usual Stokes second order coefficient (see \cite{fenton1990nonlinear}) which converges to $1/2$ for deep water waves. The third-order harmonic is given by:
\begin{align}
\eta_3 &=S_{33}\ \  k_p^2(\eta_1^3-3 \eta_1 \eta_{1H}^2),
\label{eqn:s33}
\end{align}
and the coefficient $S_{33}$ converges to $3/8$ for deep water. The detailed formulations for the approximated higher-order harmonics can be found in \cite{Walker2004}. 

We further present the spatial integral as an equivalent of a Fourier expansion \cite{taylor2024transformed} $\times 1/{\imath k}$ (i.e. $\int \cdots \text{d} x=\times  \frac{1}{\imath k}$), where $k$ is the appropriate wavenumber, $\imath = \sqrt{-1}$, and $1/{\imath}$ is effectively a Hilbert transform for sinusoidal signals. Hence the time derivative of surface elevation can be written as:
\begin{equation}
 \frac{\partial\eta}{\partial t}  = -\omega_p \eta_1-(2 \omega_p) \eta_2-(3 \omega_p) \eta_3+\cdots, 
  \label{eqn:etat}
\end{equation}
where $\omega_p$ is the spectral peak frequency of the surface elevation if we follow a slowly varying wave envelope approximation. Following a similar approach, the spatial derivative of surface elevation can be found as:
\begin{equation}
 \frac{\partial\eta}{\partial x}  = k_p \eta_1 + (2 k_p) \eta_2 + (3 k_p) \eta_3+\cdots. 
 \label{eqn:etax}
\end{equation}
Combining Equation \ref{eqn:etat} and \ref{eqn:etax}, gives  

\begin{equation}
 \frac{\partial\eta}{\partial t}  = - \frac{\omega_p}{k_p}\frac{\partial\eta}{\partial x},
 \label{eqn:etax}
\end{equation}
which can be further simplified with linear dispersion relation:
\begin{equation}
\begin{aligned}
    \frac{\partial\eta}{\partial t}  &= - \frac{g}{\omega_p} \frac{\partial\eta}{\partial x}\ \ \ \  \text{for deep water,}\\
    \frac{\partial\eta}{\partial t}  &= - \frac{g}{\omega_p}\tanh{k_pd}\frac{\partial\eta}{\partial x}\ \ \ \ \text{for finite water depth},
\end{aligned}
\end{equation}
where $d$ is the water depth.

\subsection{Independent Dataset}\label{sec:mark_data}

This independent dataset is generated through a Lagrangian simulation that solves the fully nonlinear boundary integral between the air and water interface \cite{Dold1986}. This full dataset has been generated and published independently in \cite{mcallister2023influence}. This numerical simulation stars from the fully nonlinear potential flow framework. For inviscid, rotational, and incompressible flow, a velocity potential $\phi(x , z  , t)$  satisfies Laplace’s equation:
\begin{equation}
\nabla^2 \phi=0  \text{    with    } \phi_n(x , -h  , t) \equiv 0 \text{     and   
 } \phi{}(x,-h, t) \equiv 0,
\end{equation}
where $\phi_n$ denotes the normal gradient to the bottom of the surface. A Cauchy’s integral approach is adapted to describe Lagrangian surface particle movements in time with a conformally mapped frame, which improves the computational speed when compared to the traditional boundary integral methods solved in the
physical domain. 

In this independent dataset, unidirectional focused wave groups are generated following a different spectrum is used with varying bandwidth and characteristic frequency. The JONSWAP spectrum $S_{\textrm{JONSWAP}}(f)$ used for wave initialisation following \cite{Hasselmann1973} :
\begin{equation}
S_{\textrm{JONSWAP}}(f)=G(f) \alpha_{PM} g^{2}(2 \pi)^{-4} f^{-5} \exp \left(5/4\left(\frac{f_{0}}{f}\right)^{4}\right), 
\end{equation}

where 
\begin{equation}
G(f)=\gamma^{\exp \left(\frac{\left(f-f_{0}\right)^{2}}{2 \sigma^{2} f_{0}^{2}}\right)},
\end{equation}

where $\alpha_{PM}$ is the coefficient from Pierson-Moskowitz spectra, the value of which depends on the wind conditions, $\gamma$ is the peak enhancement factor, and $\sigma$ is defined as:

\begin{equation}
\sigma=\left\{\begin{array}{ll}
0.07 & f \leq f_{0} \\
0.09 & f>f_{0}.
\end{array}\right.
\end{equation}

The Wave initialization parameters for this independent dataset are $0.71<f_p<0.88$, $1<\gamma<15$, and $0.19< Ak_p<0.37$, and a total of 149 wave groups are included in this study as an independent validation of the new breaking BC and the breaking classifier.

\subsection{Independent Experiments}\label{sec:ic_exp}
\subsubsection{Experimental set-up and wave group conditions}
The breaking waves used to validate the model presented in this study were generated in a glass-walled 2-D wave flume located at the Hydrodynamics Laboratory of the Civil and Environmental Engineering Department at Imperial College London. The flume is 27 m long and 0.3 m wide, and was filled with natural water to a fixed depth of 0.7 m throughout the experiments. A flap-type wave paddle positioned at one end of the flume generated unidirectional, dispersively focused wave groups, following the procedure outlined in \cite{Rapp1990}. This set-up allowed precise control of individual breaking events, ensuring they occurred at specific locations where an imaging system was positioned. 

This imaging system consisted of three charge-coupled device (CCD) cameras operating at 20 Hz looking sideways through the glass walls of the flume to record the breaking process. These cameras were arranged horizontally to provide a continuous spatial view of the propagating wave groups over a span of more than 4 m in the direction of wave propagation. At the opposite end of the flume, a second paddle identical to the wave generation paddle, along with a parabolic metal slope, was used to absorb and dissipate the energy of incident wave groups to minimise reflections. Reflection assessments indicated reflection coefficients consistently below $\sim$3\% for the regular wave trains tested under this configuration.

Following \cite{Cao2023}, the breaking focused wave groups were generated based upon JONSWAP-type spectra, re-formulated according to the NewWave model of \cite{Tromans1991}, which has been shown to more accurately represent the average shape of the largest crests in random sea states. In this context, breaking wave groups were defined by a discrete set of $N$ freely propagating spectral components, each with a unique frequency, amplitude and phase. The phases of individual components were tuned such that the breaking onset occurred within the field of view of the central camera in the array. 

The spectral parameters characterising the wave groups were the linear amplitude sum of all underlying components ($A$), the peak enhancement factor ($\gamma$), and the peak period ($T_p$). Specifically, for the cases presented in Figure 2 panel $(e)$, the parameters were $A=83$ mm, $\gamma=3$, and $T_p=1.2$ s. We have also compared another case with $A=90$ mm, $\gamma=2$, and $T_p=1.3$ s, where the results for the full wave breaking evolution are shown in Figure \ref{sm1}, where the surface elevation predicted through the new breaking model compares well with the experimental results within the breaking region. The full description of the functions of these parameters can be found in \cite{Cao2023} and \cite{Padilla2023}.

\subsubsection{Wave profile extraction} \label{sec:extract}
The profiles of the individual breaking waves were extracted by analysing high-resolution video images captured by the CCD cameras, which provided a spatial resolution of less than 0.001 m per pixel. Prior to extraction, the raw images were meticulously calibrated to remove lens distortion and were pre-processed to highlight the meniscus (air-water-glass contact lines), for which the contact line was used to indicate the free surface of the waves. The image processing procedures involved in this contact line enhancement included contrast enhancement, noise reduction, and edge sharpening, as outlined in \cite{Cao2024}. 

The free surface in the non-breaking regions of the wave was extracted automatically using the edge detection algorithm proposed by \cite{Cao2024}. Therein, it was termed the Continuous Maximum Gradient (CMG) method, which detects the surface by identifying the largest gradients in image pixel intensity, with an outlier removal procedure applied to filter out pixels that may be incorrectly identified as the surface points. The algorithm capitalises on the fact that the non-breaking free surface behaves as a smooth streamline, with minimal deviation between consecutive pixels representing the free surface. 

In the breaking regions, where the free surface becomes multi-valued (i.g. overturning crests) or is obscured by bubbles and surface splash-ups, the free surface was identified through free-hand outlining on a frame-by-frame basis. In these cases, the free surface was determined to be the immediate contact between the overlying air and the underlying two-phase flow. Finally, the surface elevations were obtained by converting the identified free surface in the images to real-world dimensions using pre-established relationships specific to each camera.

\bibliography{sn-bibliography.bib}% common bib file

\backmatter

\bmhead{Supplementary information}
 All code needed to replicate the given analysis can be found on github.com/menggedu/DISCOVER

\bmhead{Acknowledgements}
TT would like to acknowledge Schmidt AI in Science Postdoctoral Fellowship funded by Schmidt Science. AHC gratefully acknowledges funding from the Natural Environment Research Council [Grant NE/T000309/1

\bmhead*{Author contributions:}
Conceptualization: TT, TA, YC Formal Analysis: TT, Software: TT, YC, Funding Acquisition: TT, TA, Investigation: TT, YC, WM, TA, PT Methodology: TT, TA, YC, Data Curation: TT, YC, WM, TA, RC, AC, MM, BT, YM, PT, Supervision: TA, Writing (original): TT, Writing (review and editing): TT, YC, RC, WM, PT, MM, BT, YM, AC, TA

\section*{Declarations}
The authors declare they have no competing interests.

\bigskip

\begin{appendices}

\section{Additional Tables}\label{secA1}
\begin{table}[]
\centering
\caption{Wave initialization parameters for the numerical wave experiment, where the relevant parameters for the simulation are the liquid and gas density $\rho_w$, $\rho_a$, dynamic viscosities of liquid and gas $\mu_w$, $\mu_a$, the surface tension $\sigma$, and gravitational acceleration $g$. We have also presented the non-dimensionalised Reynolds Number and Bond number following \cite{mostert2020inertial}, where $v_w=\mu_w / \rho_w$ is the kinematic viscosity, $k_0=2 \pi / \lambda_0$ is the characteristic wavenumber and can be obtained through linear dispersion from characteristic frequency $f_0$ as $(2\pi f_0)^2 = g k_0$.}
\label{tab:param}
\begin{tabular}{@{}ll@{}}
\toprule
Parameter                                & Value                         \\ \midrule
$\rho_w$                                 & 1025 kg/m$^3$                 \\
$\rho_a$                                 & 1.225 kg/m$^3$                \\
$\mu_w$                                  & 0.00089 kg/(ms)               \\
$\mu_a$                                  & 1.74$\times$10$^{-5}$ kg/(ms) \\
$\sigma$                                 & 72 mN/m                       \\
$g$                                      & 9.81 m/s$^2$                  \\
$R e=\sqrt{g \lambda_0^3 / v_w}$         & 7000$\times$10$^3$            \\
$B o=(\rho_w - \rho_a )g / \sigma k_0^2$ & 8000                          \\ \bottomrule
\end{tabular}
\end{table}

\begin{table}[]
\centering
\caption{Model Hyperparameters for Breaking Classifier using PySR. The detailed explanation of each hyperparameter can be found in the documents \url{https://astroautomata.com/PySR/options/}}
\label{tab:my-table}
\begin{tabular}{ll}
\hline
\textbf{Hyperparameter} & \textbf{Breaking Classifier (PySR)} \\ \hline
Binary Operators        & "+" "*" "/"                         \\
Populations             & 1000                                \\
Maxsize                 & 20                                  \\
Ncyclesperiteration     & 5000                                \\
Population Size         & 33                                  \\
Loss Function           & Equation 5                          \\
Parsimony               & 0.00320                             \\ \hline
\end{tabular}%
\end{table}

\begin{table}[]
\centering
\caption{Model Hyperparameters for new Breaking BC using DISCOVER. The detailed explanation of each hyperparameter can be found in \url{https://github.com/menggedu/DISCOVER}. Other hyperparameters not listed are left as default values. }
\label{tab:hyp_discover}
\begin{tabular}{ll}
\hline
Hyperparameter           & New Breaking BC (DISCOVER) \\ \hline
Operator                 & "+" "*" "/"                     \\
Number Samples           & 150000                     \\
Batch Size               & 1500                       \\
Percentage for Migrating & 0.005                      \\ \hline
\end{tabular}
\end{table}
\section{Additional Figures}\label{secA1}

\begin{figure}
\centering
\includegraphics[width=\textwidth]{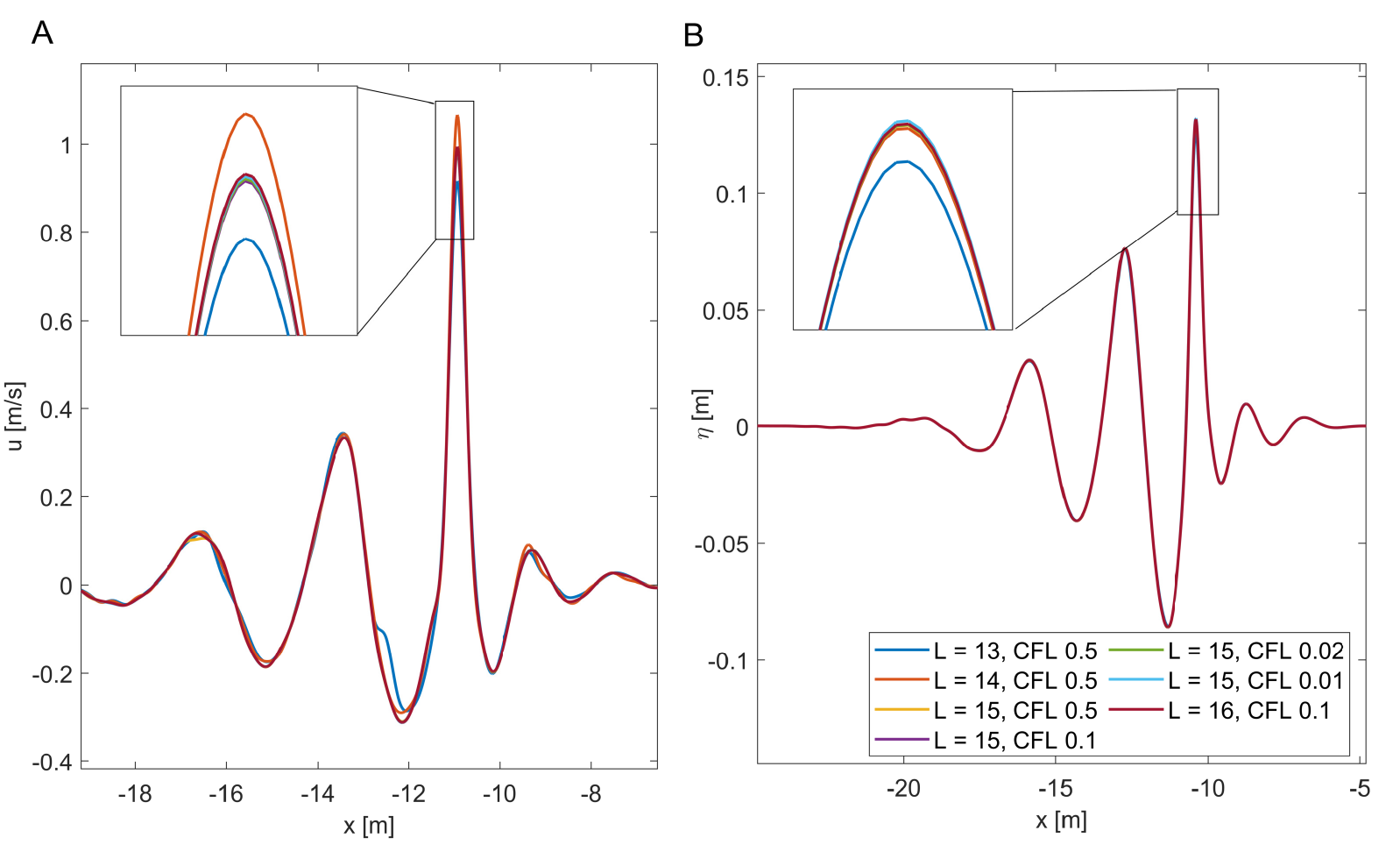}
\caption{The convergence study for panel \textbf{A} the horizontal fluid velocity at the free surface and panel \textbf{B} surface elevation itself for a breaking wave group with peak period $T_p = 1.25$, wave steepness $Ak_p = 0.24$ and bandwidth $k_w = 0.0046$m$^{-1}$. CFL shows the maximal Courant–Friedrichs–Lewy number allowed for each simulation.}
\label{conv}
\end{figure}

\begin{figure}
\centering
\includegraphics[width=0.75\textwidth]{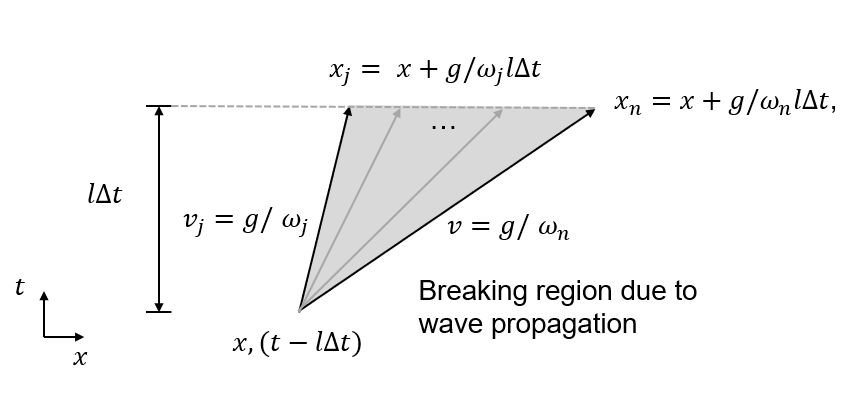}
\caption{Schematic diagram for breaking region devotion due to wave propagation}
\label{breaking_region}
\end{figure}

\begin{figure}%[tbhp]
\centering
\includegraphics[width=0.75\linewidth]{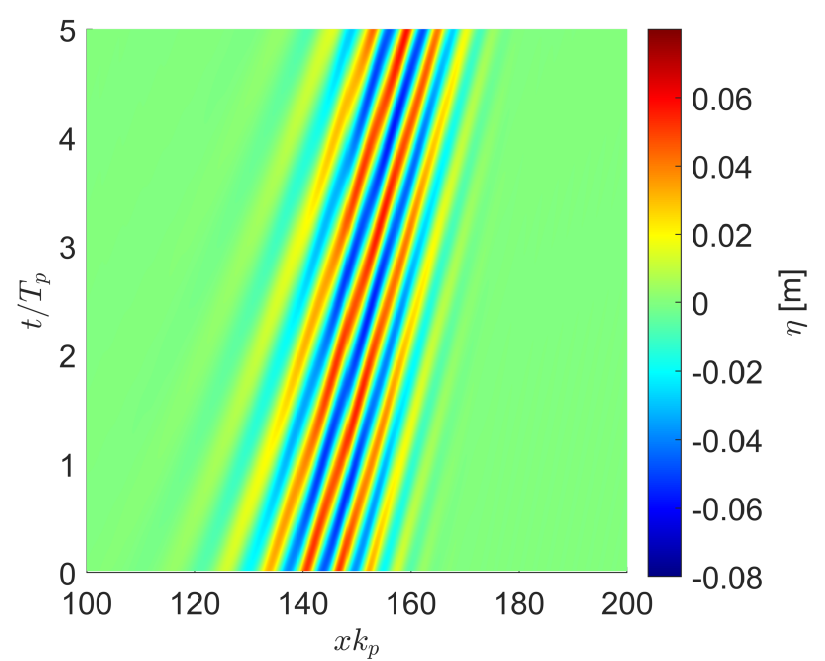}
\caption{Spatial-temporal evolution of a non-breaking wave group with new breaking BC as the only governing equation.}
\label{fig:regular}
\end{figure}

\begin{figure}
\centering
\includegraphics[width=\textwidth]{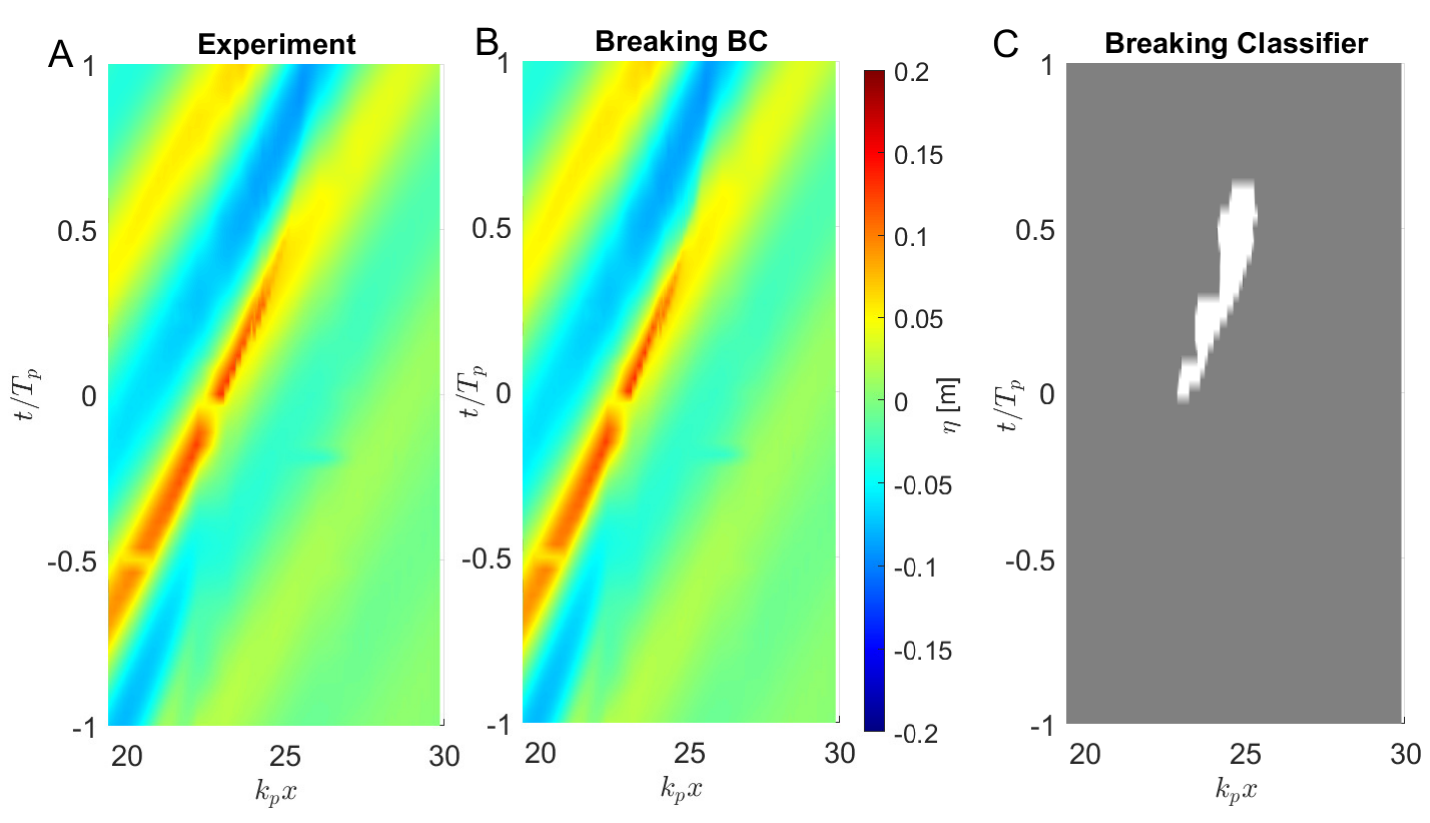}
\caption{The spatial-temporal evolution of surface elevation for a unidirectional breaking wave group for panel \textbf{A} experimental profile extracted from high-speed camera videos, and panel \textbf{B} new breaking BC being applied within the breaking region shown in panel \textbf{C} based on the evolutionary wave breaking classifier. The velocity field for classification is approximated with linear theory. The wave evolution in the rest of the non-breaking region is adapted from the experimental results to avoid edge effects during the simulation.}
\label{sm1}
\end{figure}

\end{appendices}


\end{document}